# Depth mapping of metallic nanowire polymer nanocomposites by scanning dielectric microscopy

Harishankar Balakrishnan[1], Ruben Millan-Solsona[1], Marti Checa[1], Rene Fabregas[2], Laura Fumagalli[2,3] and Gabriel Gomila[,1,4]

[1]Institut de Bioenginyeria de Catalunya (IBEC), The Barcelona Institute of Science and Technology (BIST), c/ Baldiri i Reixac 11-15, 08028, Barcelona, Spain

[2]Department of Physics and Astronomy, University of Manchester, Manchester M13 9PL, UK

[3]National Graphene Institute, University of Manchester, Manchester M13 9PL, UK4

[4]Departament d'Enginyeria Electrònica i Biomèdica, Universitat de Barcelona, C/ Martí i Franqués 1, 08028, Barcelona, Spain

**Polymer nanocomposite materials based on metallic nanowires are widely investigated as transparent and flexible electrodes or as stretchable conductors and dielectrics for biosensing. Here we show that Scanning Dielectric Microscopy (SDM) can map the depth distribution of metallic nanowires within the nanocomposites in a non-destructive way. This is achieved by a quantitative analysis of sub-surface electrostatic force microscopy measurements with finite-element numerical calculations. As an application, we determined the three-dimensional spatial distribution of ~ 50 nm diameter silver nanowires in ~ 100–250 nm thick gelatin films. The characterization is done both under dry ambient conditions, where gelatin shows a relatively low dielectric constant, $\varepsilon r$ ~ 5, and under humid ambient conditions, where its dielectric constant increases up to $\varepsilon r$ ~ 14. The present results show that SDM can be a valuable non-destructive subsurface characterization technique for nanowire-based nanocomposite materials, which can contribute to the optimization of these materials for applications in fields such as wearable electronics, solar cell technologies or printable electronics.**

## Introduction

Polymer nanocomposite materials made with metallic nanowires as fillers are being investigated due to their enhanced optical, electromagnetic, electrical, mechanical and thermal properties.[1] Due to the high aspect-ratio of metallic nanowires, the addition of small amounts of them can drastically modify the physical properties of the polymeric matrices such as conductivity,[2] dielectric constant,[3] electromagnetic shielding,[4] thermal conduction,[5] etc. Other relevant physical properties, like transparency, flexibility or stretchability, remain almost unaffected. This fact enables developing novel materials with combined physical properties for use as transparent electrodes for solar cell applications,[6–8] as flexible electrodes for wearable electronics,[9,10] and as high dielectric constant stretchable dielectrics for highly sensitive pressure sensors.[11] Metallic nanowire nanocomposite materials are, in addition, solution processable[12,13] and printable.[14,15]

The macroscopic physical properties of metallic nanowire nanocomposites depend on the dimensions of the nanowires (diameter and length), their concentration (in % weight of the composite) and their spatial distribution within the polymeric matrix. The information on the 2D spatial distribution of the nanowires on intact samples can be obtained by optical microscopy on transparent samples (metallic nanowires are usually detectable in transparent samples although with a spatial resolution limited by the diffraction limit of light), scanning electron microscopy (although with some limitations due to the insulating nature of the polymer matrices and the need to use low electron doses to prevent sample damage, which can affect the spatial resolution and depth sensitivity) and atomic force microscopy (since usually the buried nanowires induce small protrusions on the upper surface). To gain information on the 3D spatial distribution of the nanowires one should resort to nanoscale subsurface and tomographic techniques. The current gold standard nanoscale tomographic techniques in materials science are based on electron and X-ray microscopies.[16–18] To these techniques we could also add confocal optical microscopy, with a spatial resolution in the three spatial directions limited by the diffraction limit of light. However, these techniques have not been much applied to metallic nanowire polymer nanocomposites yet.[19] Alternative nanoscale tomographic techniques compatible with the properties of polymeric materials and realization of measurements under ambient conditions have also been investigated. Among them, we find those based on sub-surface sensitive Scanning Probe Microscopy (SPM).[20,21] Techniques such as Scanning Near Field Ultrasound Holography,[22,23] Mode Synthesizing Atomic Force Microscopy (AFM),[24] Multimodal AFM,[25] Amplitude Modulated AFM,[26,27] Electrostatic Force Microscopy,[28,29] Scanning Microwave Impedance Microscopy,[30–34] or Scanning Near Field Optical Microscopy[35] have demonstrated the capability to image buried nanostructures within polymeric materials.

Electrostatic Force Microscopy (EFM) is among the subsurface SPM techniques that has progressed more towards its implementation as a nanotomographic technique compatible with polymeric materials.[36–41] In EFM a voltage bias is applied between the conductive probe of an AFM system and the sample, and the electric force acting on the probe is measured while the tip is scanned along the sample surface. Due to the long-range nature of the electric forces, EFM can sense the presence of nanoscale objects buried below the surface. Examples of applications include the imaging of carbon nanotubes (CNT),[29,36,42–46] graphene networks[47] and nanoparticles[48] buried in polymer nanocomposites. To obtain nanotomographic information a quantitative analysis of the sub-surface EFM images is required, as has been shown in ref. 29 and 36, where the depth distribution of CNTs in polymer films has been derived. Recently, in ref. 38, the subsurface EFM images of water-filled nanochannels buried into a dielectric material were combined with the quantitative analysis of the tip–sample capacitive interaction using finite-element numerical calculations. This approach, referred to as Scanning Dielectric Microscopy (SDM),[49,50] allowed one to precisely determine the dielectric constant of water confined in single nanochannels, from tens-of-nm thick highly polarized bulk water down to a few molecular layers of low-polarized water buried under a thick dielectric.

Here, we applied Scanning Dielectric Microscopy to metallic nanowire nanocomposites. We demonstrate that by quantitatively analysing sub-surface EFM measurements by means of finite element numerical calculations, the depth distribution of the metallic nanowires can be determined in a non-destructive way with nanoscale spatial resolution. As compared to similar studies performed earlier with CNT polymer nanocomposites,[29,36] we had to deal here with additional challenges imposed by the much larger dimensions of the metallic nanowires that induced a non-planar surface of the nanocomposites and made relevant the capacitive coupling between them. As an application we have considered the case of a silver nanowire/gelatin (AgNW/gelatin) nanocomposite. This nanocomposite constitutes an example of a nanocomposite made from bio-renewable resources.[51] Besides the intrinsic interest in the development of biosensors,[52] this nanocomposite offers also the possibility of investigating easily the effect of varying the matrix dielectric constant, since gelatin passes from $\varepsilon_r \sim 4$ under dry ambient conditions to $\varepsilon_r \sim 15$ under humid ambient conditions.[53] The demonstration that nanotomographic SDM information can be obtained on nanocomposites with high dielectric constant matrices is another important difference from earlier nanotomographic EFM works, which systematically considered polymeric materials with very low dielectric constants ($\varepsilon_r \sim 2$–$3$) that offered optimal conditions for sub-surface EFM characterization.

**Results**

AgNWs/gelatin nanocomposites have been prepared by first drop casting $\sim$50 nm diameter AgNWs on top of a highly doped silicon substrate and then spin-coating gelatin on top of them and leaving the sample to dry. Fig. 1 shows a 60 μm × 60 μm topographic AFM image of one of the samples analysed.

A scratch has been made on purpose to measure the sample thickness, giving in this case a thickness $t_m \sim 254$ nm ± 1 nm (see the cross-section profile in Fig. 1b and histogram analysis in Fig. 1c). The AgNWs within the gelatin film are clearly visible in the topographic AFM image, where they appear as small protrusions on the, otherwise very flat, gelatin surface (the rms roughness of gelatin is $\sim$ 1 nm). The protrusions have heights in the range of $\sim$ 2–15 nm, most of them in the lower bound range, and widths in the range of $\sim$ 200–1000 nm (see the inset of Fig. 1c where the height versus width of some characteristic protrusions is shown). The dimensions of the protrusions have nothing to do with the dimensions of the AgNWs, which are $\sim$ 50 nm in diameter. A rough linear relation between the height and the width of the protrusions seems to exist, but this fact does not seem to be apparently correlated with the depth position of the nanowires. The AFM topographic image, then, reveals the presence of the nanowires buried in the gelatin film, and can provide an overview of its 2D distribution. However, it does not provide information about the depth at which the nanowires are buried. In particular, it does not give information on the separation existing between the crossing nanowires, which is a parameter of relevance to determine the overall macroscopic electrical properties of the nanocomposite.

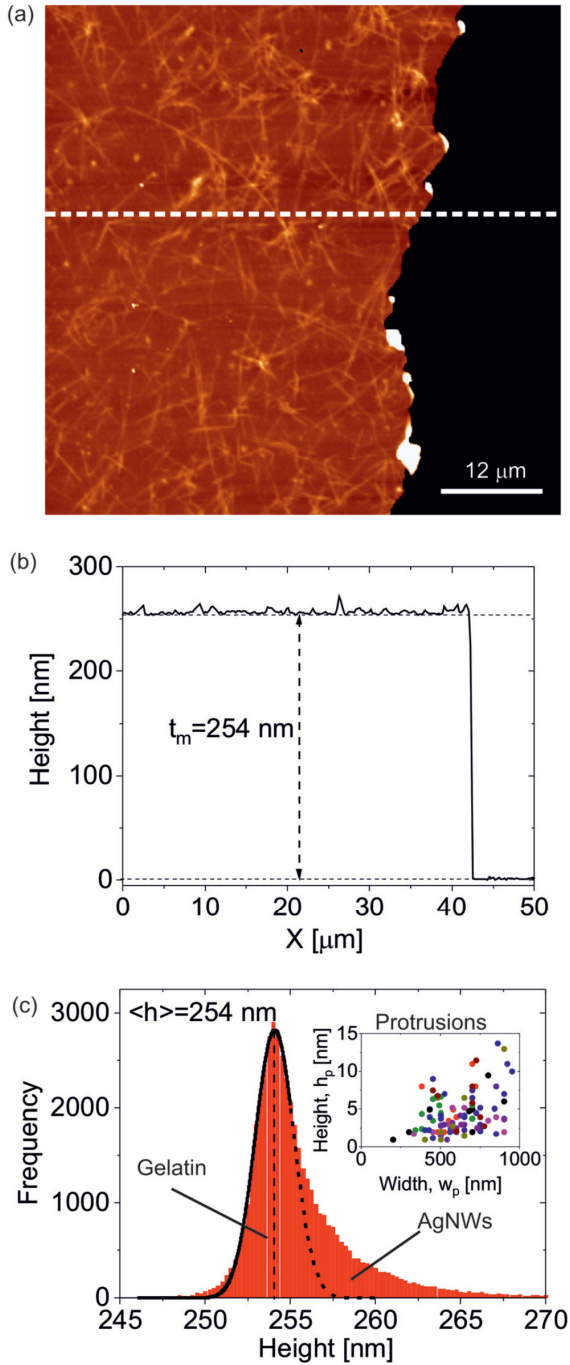

Figure 1. (a) Large scale AFM topographic image (60 × 60 μm²) of a AgNW/gelatin film ~ 254 nm thick containing ~ 50 nm diameter silver nanowires on a highly doped silicon substrate. The gelatin film has been scratched on purpose to determine its thickness. (b) Cross-section topographic profile along the dashed line in (a). The small protrusions correspond to the buried nanowires. (c) Histogram analysis of the distribution of heights of the gelatin film. The continuous line is a Gaussian fit to half of the distribution, representing the bare gelatin parts (the dashed line is an extrapolation of the Gaussian fit). The heights outside the Gaussian distribution correspond to the protrusions due to the buried nanowires. Inset: height vs. width of some representative protrusions present in (a).

In order to obtain information on the depth distribution of the nanowires, EFM measurements have been performed. Here we used the force volume acquisition mode presented recently,[54] in which full ac electric force approach curves are acquired at each pixel of the image. From the force volume data, EFM images can be derived by post-processing in any desired mode and at any desired imaging distance.[54] Fig. 2a–e and f–j show, respectively, higher resolution AFM topographic and constant height EFM images corresponding to selected regions of the sample in Fig. 1a. All the EFM images correspond to a tip-gelatin distance $z = 36$ nm ($z = 290$ nm with respect to the substrate). The EFM images are expressed, as usual,[49] in terms of the capacitance gradient, $dC/dz$, which is related to the ac electric force amplitude at the $2\omega$ harmonic by $F_{2\omega} = 1/4 \cdot dC/dz \cdot v_0^2$, here $v_0$ is the amplitude of the applied ac voltage. The EFM images, like the topographic AFM ones, clearly reveal the presence of the buried nanowires. At the first sight a clear correlation exists between the AFM topographic and EFM images. This correlation is not related to the topographic crosstalk present in lift-mode EFM images.[55] The correlation exists due to the non-planar nature of the sample surface, which indicates that the constant height EFM images display higher electric forces at the locations of the topographic protrusions, since they are closer to the tip. The measured electric forces, however, not only contain information on the topographic protrusions, but also provide information on the electric polarization of the buried nanowires. This fact can be seen directly in a few cases in which the nanowires are only visible in the EFM images and not in the AFM topographic images (for instance the nanowire highlighted with an arrow in Fig. 2g).

In order to disentangle the contributions of the topographic protrusions and of the polarization of the buried nanowires to the electric force, we have considered the model shown in Fig. 2k (not to scale). The model includes both a surface protrusion and a buried nanowire. Note that in earlier studies on CNT nanocomposites,[29,36] it was not necessary to include the surface protrusions since CNTs are much smaller than AgNWs and hence they perturb much less the surface of the nanocomposites. Fig. 2l shows an example of a calculated electric potential distribution obtained with the model in Fig. 2k, which shows the locality of the electric interaction. In Fig. 2l the dashed lines at the centre of the image represent the ends of the protrusion centred on the nanowire, which otherwise is imperceptible at the scale of the representation. Fig. 2m shows the numerically calculated dependence of the capacitance gradient contrast at the center of the nanowire, $\Delta C'_0$, as a function of the depth position of the nanowire, d, for the tip at $z = 36$ nm from the flat part of the sample and for different heights of the surface protrusions, $h_p$. Since the width of the protrusions, $w_p$, plays a minor role and it has been kept fixed to a representative value $w_p = 620$ nm. For a given protrusion height, the graph shows that the capacitance gradient contrast $\Delta C'_0$ shows a strong dependence on the depth position of the nanowire, thus enabling to determine the depth from the measured $\Delta C'_0$.[29,36] The graph also shows that the protrusion height induces a vertical shift of the capacitance gradient contrast $\Delta C'_0$ curves. This latter result explains why there is a correlation between the AFM topographic and EFM images, and, indicates that the presence of the protrusions needs to be accounted for explicitly in the quantitative analysis. Finally, we note that, in the present case, the EFM measurements are sensitive to the depth position of the nanowires in the whole thickness of the gelatin film, since the measuring noise is ~1 zF nm$^{-1}$ and the minimum contrast generated by a buried nanowire (when at the bottom of the gelatin film) is ~5 zF nm$^{-1}$ (see Fig. 2m). Based on these results, to determine the depth position of the nanowires inside the gelatin film we have proceeded as follows. At a given position on top of a nanowire, we determined the height of the protrusion, $h_p$, from the topographic image (the width is kept the same for all protrusions, $w_p = 620$ nm, as mentioned above). Then, for the given protrusion height, we calculated theoretical $dC/dz$ approach curves with the model in Fig. 2k for different depths, $d$, of the nanowire and fitted them with the experimental $dC/dz$ approach curve acquired at the given position, with the depth, $d$, as the single fitting parameter. This process is repeated at all points along the nanowires. The rest of the parameters of the model are determined as follows: the thickness of the gelatin film ($t_m = 254$ nm) is obtained from the large-scale topographic image (Fig. 1a), the tip radius ($R = 87 \pm 2$ nm), half cone angle ($\theta = 28.3 \pm 0.4°$), both consistent with manufacturer specifications, and capacitance gradient offset ($C'_{offset} = 107 \pm 2$ zF nm−1) from the $dC/dz$ approach curves acquired on the bare substrate (see Fig.S1 in SI), and the relative dielectric constant of gelatin ($\varepsilon_m = 13.8 \pm 0.3$) from the $dC/dz$ approach curves acquired in gelatin regions that do not contain any nanowire (see Fig.S1 in SI).

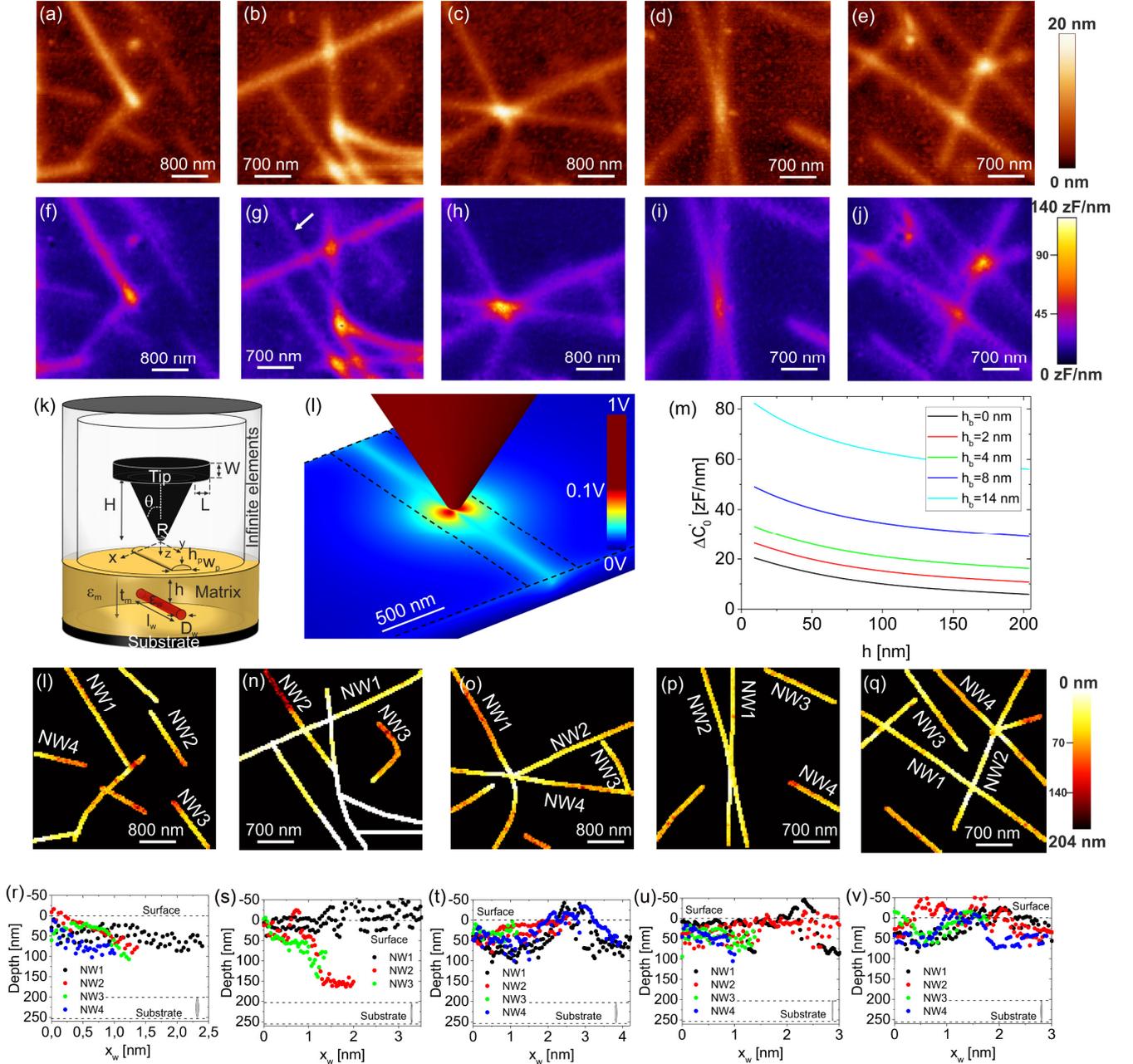

Figure 2. (a)–(e) Topographic AFM and (f)–(j) capacitance gradient constant height EFM images ($z = 36$ nm) obtained in five selected regions of Fig. 1a. Experimental parameters: probe spring constant $k = 0.57$ N m$^{-1}$, resonance frequency $f_0 = 17$ kHz, voltage frequency $f_{el} = 2$ kHz, and voltage amplitude $v_0 = 4$ V. (k) Schematic representation of the buried nanowire model with surface protrusion used to quantify the EFM data. (l) Example of a calculated electric potential distribution. Parameters of the calculations: Tip–sample distance $z = 36$ nm, tip radius, $R = 87$ nm, half cone angle, $\theta = 28°$, gelatin thickness, $t_m = 254$, gelatin dielectric constant, $\varepsilon_m = 14$, nanowire diameter $D_w = 50$ nm, nanowire length, $l_w = 8$ μm, nanowire dielectric constant $\varepsilon_w = 105$ (metallic), and width of the protrusion $w_p = 620$ nm. (m) Calculated maximum capacitance gradient contrast, $\Delta C'_0$, at the centre of the nanowire as a function of the depth position of the nanowire for different heights of the surface protrusions. The same parameters as in (l). (n)–(r) Depth maps of the nanowires corresponding to the five regions analyzed, determined with the model in (k). The depths have been determined by fitting calculated full approach curves at each pixel on top of the nanowires. (s)–(w) Depth profiles along some of the nanowires present in (n)–(r). The origin of the negative depths is explained in the text.

For the nanowires we considered a diameter $D_w = 50$ nm, as a representative value according to manufacturer specifications and to our own earlier characterization[56] (see Fig. S2 in SI for the effect of the nanowire diameter in the extracted depths). For the length of the nanowire, we took $l_w = 8$ μm (see Fig. S3 in SI for the effect of the nanowire length on the results). Finally, we assumed a very large dielectric constant for the nanowire ($\varepsilon_w = 105$), which corresponds to the metallic limit. Fig. 2n–r show the maps of the depth position of the nanowires within the gelatin matrix determined in this way corresponding to the five regions analyzed in Fig. 2f–j. Fig. 2s–w show depth profiles taken along some representative nanowires present in Fig. 2n–r. The first relevant aspect noted is that most nanowires seem to be located close to the gelatin top surface (depths ~ 10–75 nm), rather than at the bottom surface. This result is remarkable since when the sample was prepared the nanowires were initially spread on the bottom substrate. Another relevant aspect of Fig. 2n–r and s–w is that the depth maps reveal the inclination of some of the nanowires (e.g. NW2 and NW3 in Fig. 2s and t, respectively). Detecting the inclination of the nanowires with a model like the one in Fig. 2k, which considers a non-inclined nanowire, is possible because of the very local nature of the tip-nanowire interaction, as it is demonstrated in Fig. S4 in SI. Finally, from the depth maps we can identify for the crossing nanowires which one is located on the top and which one at the bottom, which is not apparent from the topographic or EFM images. For instance, in Fig. 2o NW2 crosses below NW1, while in Fig. 2p NW2 crosses above NW4.

At the crossing points between different nanowires, the depths tend to show smaller values than for the rest of the nanowire (and even sometimes the values are unphysical, e.g. negative). There are two possible effects responsible for this fact, namely, the fact that the surface protrusion of crossing nanowires can be locally much wider than the one assumed for the single nanowire model in Fig. 2k, and the fact that multiple nanowires can contribute to the measured electric force (only a single nanowire is included in the model of Fig. 2k). Both effects would lead to larger electric forces, as observed experimentally. The way how multiple nanowires contribute to the measured electric force in SDM merits some comments, since it is not trivial due to the relatively large size of the nanowires and to the strong dependence of the electrostatic force on the depth position.[40] Again, the situation is very different from what would occur for CNTs in nanocomposites,[29,36] since they are much smaller in diameter. To analyze the electric force generated by crossing nanowires we have considered a model like the one in Fig. 2k but with two nanowires crossing at an angle, α (see the inset in Fig. 3a). The nanowires are assumed to be parallel to the substrate, and, for simplicity, no surface protrusion is considered ($h_p = 0$ nm). Fig. 3a shows a calculated constant height $\Delta C'$ EFM image ($z = 36$ nm) for this model for a crossing angle α = 45° and the two nanowires in close contact (interwire vertical separation, from edge to edge, $\Delta z_w = 0$ nm). Fig. 3b and c show the $\Delta C'$ profiles taken along the transversal and longitudinal directions of the top nanowire, respectively (dashed lines in Fig. 3a), for different interwire vertical separations $\Delta z_w = 0$ nm (black line), 20 nm (grey line) and 100 nm (red line), where only the bottom nanowire is displaced. For comparison, Fig. 3b and c also show the $\Delta C'$ profiles corresponding to the two nanowires when they are alone (dashed black and green lines respectively). The most relevant feature of the calculations is that the bottom nanowire contributes to the total calculated capacitance gradient, ΔC', mainly in the crossing region and only by a small amount (~ 1–1.5 zF nm$^{-1}$). This contribution is much smaller than the one corresponding to an isolated nanowire at the same depth (which ranges from ~ 2 to 8 zF nm$^{-1}$). The reason is that the top nanowire screens the electric field below it at the crossing point preventing the polarization of the bottom nanowire in this region. The actual contribution to the total capacitive gradient of the bottom nanowire can be seen in Fig. 3d by comparing the total capacitance gradient contrast at the centre of the top nanowire, $\Delta C'_0$, with the signal due to the top nanowire when alone (the dashed line), as a function of the interwire separation and crossing angle (continuous lines). As said, the contribution of the bottom nanowire is just ~ 1–1.5 zF·nm$^{-1}$ over a total of 16–18 zF·nm$^{-1}$. This small contribution alone cannot explain the relatively large variations of the electric forces observed at the crossing points of different nanowires, from what we conclude that most of the excess electric force should be due to a local relatively large variation of the width of the topography of the surface due to the crossing wires.

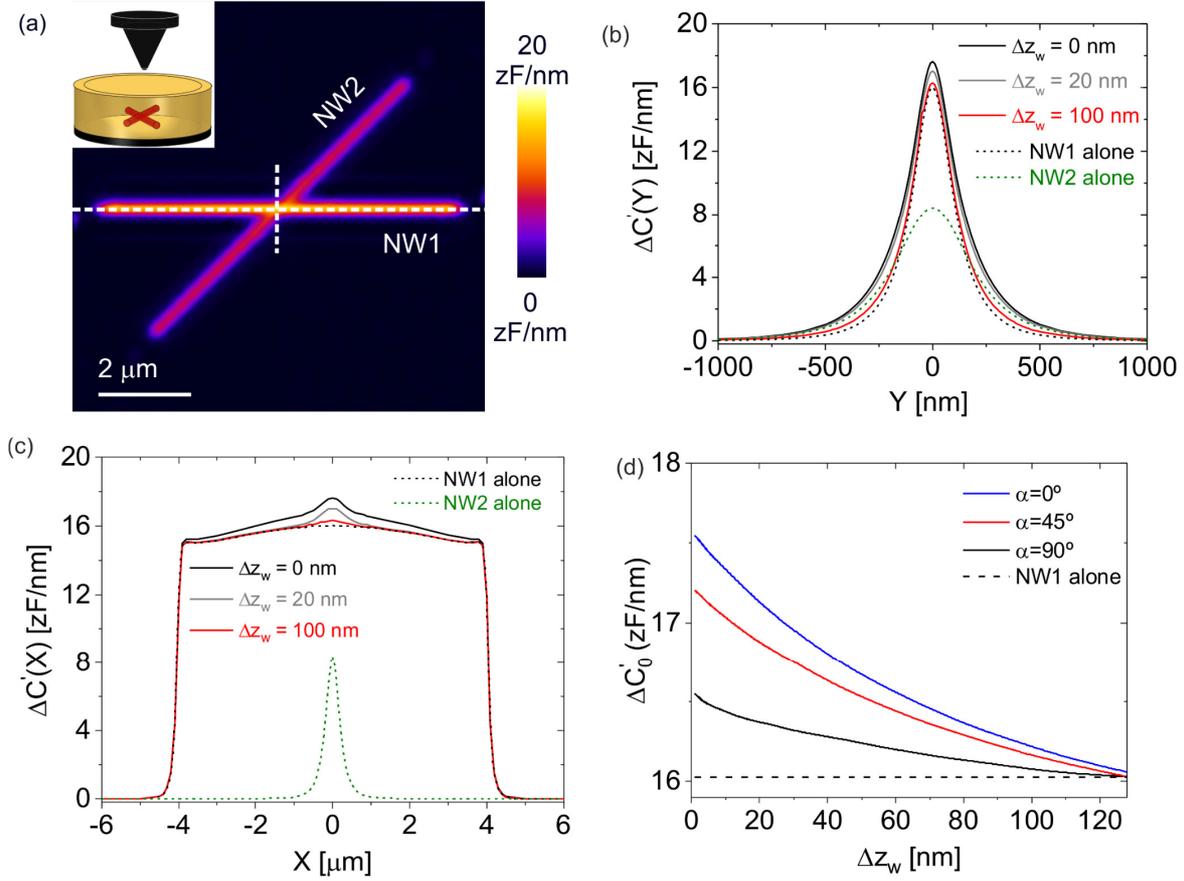

Figure 3. (a) Numerically calculated constant height EFM image corresponding to two nanowires parallel to the substrate crossing at an angle α = 45° and with a vertical separation $\Delta z_w$ = 0 nm (see Inset). Parameters of the calculations: the same as in Fig. 2, $z$ = 36 nm, $D_{w1} = D_{w2}$ = 50 nm, $l_{w1} = l_{w2}$ = 8 μm, $\varepsilon_{w1} = \varepsilon_{w2}$ = 105 (metallic), $d_1$ = 10 nm, $d_2$ = 60 nm, tm = 254 nm, $\varepsilon_m$ = 13.8, $R$ = 87 nm, θ = 28°, $H$ = 12.5 μm, $W$ = 3 μm, $L$ = 3 μm. (b) and (c) (continuous lines) ΔC' profiles across and along the top nanowire (dashed lines in (a)), respectively, for three different depths of the bottom nanowire $d_2$ = 60 nm, 80 nm and 160 nm, corresponding to vertical interwire separations $\Delta z_w$ = 0 nm, 20 nm and 100 nm, respectively. For comparison, we have also plotted the profiles corresponding to single nanowires located at the positions of nanowires 1 and 2 (dashed black and green lines, respectively). (d) Capacitance gradient contrast $\Delta C'_0$ at the centre of the top nanowire (X = Y = 0) as a function of the vertical separation between nanowires, $\Delta z_w$, for three crossing angles (α = 0°, 45° and 90°). For comparison the value corresponding to a single nanowire at the position of the top nanowire is shown (the dashed line).

The depth mapping capabilities of SDM have also been demonstrated on AgNW/gelatin nanocomposites under dry ambient conditions, where the dielectric constant of gelatin lowers down to $\varepsilon_r \sim 4.8$, as compared to $\varepsilon_r \sim 14$ under humid ambient conditions (see Fig.S5 in SI). Fig. 4a shows an AFM topographic image of one of the samples analyzed. The thickness of the nanocomposite film is here $t_m \sim 98$ nm (see the topographic cross-section profile in Fig. 4b). Like the sample analyzed in Fig. 2, the surface of the gelatin is fairly flat (rms roughness ~ 1 nm), and it only shows some very small topographic protrusions induced by the buried nanowires (a few nanometers high and hundreds of nanometers wide). Fig. 4c shows a constant height EFM image corresponding to a tip–substrate distance $z = 130$ nm (i.e. at ~ 32 nm from the gelatin surface). Fig. 4d shows a capacitance gradient cross-section profile along the dashed line in Fig. 4c. The buried nanowires are again clearly seen in the EFM image. Fig. 4e shows the corresponding depth map, while Fig. 4f shows a cross-section profile taken along the dashed line in Fig. 4e. In Fig. 4e, we have also drawn the cross-section of the nanowires (which due to the scales of the graphical representation appear as stretched ovals). The parameters used in the quantitative analysis to derive the depths are indicated in the caption of Fig. 4 (see also Fig.S5 in SI). Finally, Fig. 4g shows the depth profiles taken along some of the nanowires present in Fig. 4e. In Fig. 4g the dashed lines indicate the values of the depth for which the nanowire will be lying at the surface and at the bottom substrate, respectively. In the present case the nanowires seem to lie close to the substrate, and mostly parallel to it. In the region where the nanowires cross each other, much lower depths are obtained, for the reasons explained above.

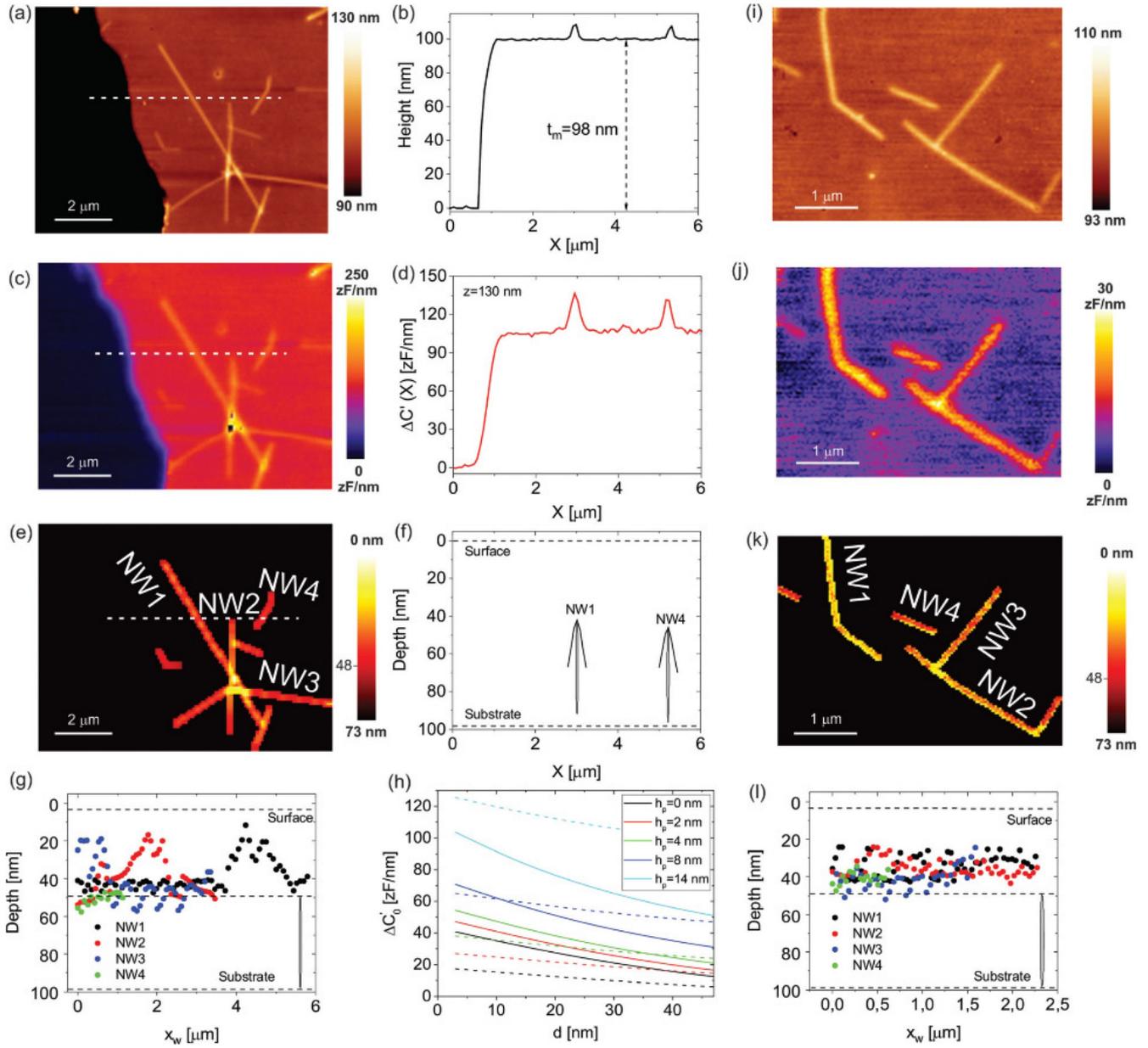

Figure 4. (a) Topographic image of a gelatin/AgNW nanocomposite under dry ambient conditions. A scratch has been made on the gelatin nanocomposite to determine its thickness. Experimental parameters: $k$ = 0.39 N/m, $f_0$ = 17 kHz, $f_{el}$ = 2kHz and $v_0$ = 4 V. (b) Topography cross-section profile along the dashed line in (a). (c) Constant height EFM image reconstructed at a distance $z$ = 130 nm with respect the substrate. The black spots in the image corresponds to regions whose topography is higher than the distance at which the image is reconstructed. (d) Capacitance gradient contrast profile, $\Delta C'(X)$, along the dashed line in (c). (e) Depth map of the nanowires present in the region analyzed in (a). Parameters used in the quantitative analysis: $R$ = 136 nm, $\theta$ = 29.3º, $t_m$ = 98 nm, $\varepsilon_m$ =4.8, $w_p$ =220 nm, $D_w$ = 50 nm, $l_w$ = 8 µm, $\varepsilon_w$ = $10^5$ (metallic). (f) Depth cross-section profile along the dashed line in (e). The cross-section of the nanowires has been drawn to the scale of the figure. (g) Depth profiles along some selected nanowires present in (e). (h) Calculated contrast at the center position of the nanowire for $z$ = 32 nm as a function of the depth position for different heights of the surface protrusions. Continuous and dashed lines are for $\varepsilon_m$ = 4.8 and 13.8, respectively. The rest of parameters are the same as in (e). (i)-(l) Same as (a), (c),(e) and (g) but for a different region of the sample.

The crossing nanowires do not seem to be located at different depths, which could indicate that the nanowires sort of wrap one around the other at the crossing points. In the present case, the contribution of the protrusions to the measured capacitance gradient contrast, $\Delta C'$, is less significant (due to the smaller dielectric constant of gelatin) and the sensitivity to the depth is higher. We show it explicitly in Fig. 4h where we plot the variation of capacitance gradient contrast at the center of a nanowire, $\Delta C'_0$, as a function of the depth position of the nanowire, h, for different heights of the gelatin surface protrusions, hp. The continuous lines correspond to the actual dielectric constant, $\varepsilon_m = 4.8$, while the dashed lines correspond to the values that would be expected under humid conditions, where $\varepsilon_m = 13.8$. When the height of the protrusions increases, the capacitance gradient curves shift up vertically like before, but the shifts are much smaller for the dielectric constant $\varepsilon_m = 4.8$ than those for $\varepsilon_m = 13.8$. This fact shows that the protrusions contribute much less to the capacitance gradient values under dry conditions, as advanced above. Moreover, for a given protrusion height, the capacitance gradient shows a steeper dependence on the depth for $\varepsilon_m = 4.8$ than for $\varepsilon_m = 13.8$, meaning a higher sensitivity to the depth position of the nanowires. Fig. 4i–l show a similar analysis to the one performed in Fig. 4a, c, e and g but corresponding to another region of the sample. This analysis confirms the main findings described before.

**Discussion**

We have shown that the depth distribution of silver nanowires in gelatin nanocomposites can be determined in a nondestructive way by means of scanning dielectric microscopy. As compared to earlier depth distribution analysis performed by EFM on CNT polymer nanocomposites,[29,36] some important differences are worth being highlighted. In earlier studies the sample surface was very flat, such that the CNTs were not identifiable in the topographic images. This is probably due to the very small dimensions of CNTs (~ 2–5 nm diameter) as compared to AgNWs (~ 50 nm). The possibility to address non-flat surface nanocomposite materials constitutes an important advantage of the present approach, since in many applications the nanocomposite surface roughness due to the buried nanowires cannot be avoided.[9] On the other side, we have analyzed here the effect that the matrix dielectric constant can have on the depth mapping capabilities of SDM (or EFM). We have seen that for relatively large dielectric constant materials (e.g. $\varepsilon_m > 10$) the depth mapping becomes more complex since the electric field is more strongly screened within the material, and also because any surface protuberance induces a larger parasitic electric force variation, not related to the polarization of the buried nanowire. We have shown that depth mapping is still possible under these conditions if properly implemented, as we have done here. This result enlarges the range of materials whose sub-surface nanoscale properties can be probed, including materials whose dielectric constant is humidity responsive, like gelatin. Earlier studies were almost exclusively carried out in low dielectric constant materials with $\varepsilon_m \sim 2$, in which electric field screening or topographic effects are much less relevant, as we have shown from the measurements performed under dry conditions where gelatin shows $\varepsilon_m \sim 5$.

Here, we have also analyzed the effect that crossing nanowires can have on the measured SDM signal, an aspect that was not considered in CNT nanocomposite studies, since only single buried CNT models were considered.[29,36] We have shown theoretically that the electric force at the crossing points is not the addition of the electric forces due to the two nanowires isolated, but much smaller. In fact, we have seen that at the crossing points the bottom nanowire contribution is almost negligible, while far from the crossing point it is easily detectable. We have explained this non-intuitive fact as due to a local electric screening of the top nanowire on the bottom nanowire at the crossing point. When the vertical separation of the nanowires is smaller than their diameter, a second effect could also play a role, namely, the capacitive coupling between the nanowires.[40] Due to the capacitive coupling some of the electrostatic energy provided to the system is stored between the nanowires, and hence the electric force acting on the tip is reduced with respect to the simple addition of electric forces due to each separate nanowire. In the present study, we could not identify clearly this effect, since it was masked by a much larger effect due to the local surface topography modification induced by the crossing wires. Still, the possibility to detect the local capacitive coupling between the nanowires in a network should be kept in mind, since it could give an indication of the percolation level of the network, a parameter of maximum relevance to determine the overall electric macroscopic properties of the nanowire nanocomposite.[2–4]

Concerning the achievable lateral spatial resolution, it depends on several factors, including the tip radius, half cone angle, polymer thickness and dielectric constant, and depth position of the nanowires, as we have shown in a dedicated theoretical study.[40] Based on that study, the achievable lateral spatial resolution in a sample like the one in Fig. 2 is below 50 nm for shallow buried nanowires. For nanowires located at the bottom of the film (at around ∼ 200 nm depth) the spatial resolution decreases to ∼ 200 nm due to the broadening of the electric field lines. The spatial resolution has been determined from finite element numerical calculations performed with a model like the one in Fig. 2k but involving two buried parallel nanowires. The spatial resolution is defined as the edge to edge separation of the nanowires for which the contrast between the middle point between the nanowires and the centre of one of the nanowires is at least twice the instrumental noise. Concerning the vertical spatial resolution, since nanowires are extended objects, it is basically limited by the depth uncertainty, $\delta d$, by which the depth of the nanowires can be determined. The depth accuracy is given by $\underline{\delta d} = \delta C'_{noise}/|\partial \varDelta C'/\partial d|$ where $|\partial \varDelta C'/\partial d|$ is the sensitivity of the capacitance gradient contrast to variations in the depth position of the nanowire and $\delta C'_{noise}$ the instrumental capacitance gradient noise (here 1 zF nm−1). By using the data in Fig. 2m it can be shown that the depth uncertainty for the sample in Fig. 2 is 2–6 nm for a nanowire located close to the film surface and 12–32 nm for a nanowire at the bottom of the film (the ranges depend on the height of the protrusions). Similarly, by using the data in Fig. 4h it can be shown that the depth accuracy for the sample in Fig. 4 is just 0.5–1 nm for a nanowire close to the surface of the film and 1–3 nm for a nanowire at the bottom of the film. By improving the capacitance gradient instrumental noise of SDM in the force volume acquisition mode to match the values reported for the conventional imaging mode (0.1 zF nm−1),[49] nanometric depth accuracies are expected to be generally achievable.

Finally, we highlight the main advantages of the proposed approach with respect to alternative tomographic approaches based on optical or electron microscopy. Concerning optical microscopy, the main advantage is mainly the higher achievable lateral and vertical spatial resolutions, and the possibility to apply it to non-transparent samples. Concerning electron microscopy, the main advantage is that its performance is not severely affected by the fact of applying it to samples under ambient, and even, liquid conditions, and its gentle non-damaging interaction with soft samples. Environmental and air scanning electron microscopes have been developed, but they show a reduced performance with respect to vacuum operated electron microscopes, which already find limitations when applied to nanoobjects buried in polymeric matrices.[57] We note that like electron microscopy, the presented approach is not seriously affected by the overall thickness of the nanocomposite film,[50] although it is only sensitive to the first hundreds of nanometres of the specimen. Concerning imaging speed, scanning probe microscopy techniques are traditionally considered slow imaging techniques. However, the present commercially available atomic force microscopes can obtain images in just seconds. We have argued in ref. 54 that in these systems the accurate SDM images needed for nanotomographic imaging could be acquired in less than one minute.

**Conclusions**

In summary, we have shown that the depth distribution mapping of metallic nanowires in polymer nanocomposites can be conducted in a non-destructive way by means of scanning dielectric microscopy. To achieve this we have overcome the challenges imposed by the non-planar surface of these nanocomposites and the eventually large dielectric constant of the matrix. The depth maps provide information on the vertical distribution of the nanowires, and they can also provide useful information to investigate the percolation level of the nanocomposites, which is of utmost relevance in determining the overall macroscopic electrical properties of the composite materials. The present results are expected to contribute to the optimization of the properties of metallic nanowire nanocomposites and to push forward their application in solar cell technologies and wearable electronics, among others.

## Methods

**AgNW/gelatin nanocomposite preparation.** To prepare the gelatin nanocomposites we used a highly doped silicon substrate. Prior to its use, the substrate was cleaned with acetone (PanReac, Applichem) and 2-Propanol (Sigma Aldrich, ACS reagent) in an ultrasonic water bath and thoroughly dried. Silver nanowires (A50 Research Grade, 0.5 g dissolved in 50 mL IsoPropyl Alcohol, IPA) were obtained from Novarials (Novawire-Ag-A50). The nanowires (AgNWs) were further diluted with IPA to produce a relatively sparse network of nanowires in the nanocomposites. The AgNWs were drop casted on the Si substrate and let to dry in a vacuum chamber for 3 hours. Gelatin from Porcine Skin – Type A was procured from Sigma Aldrich. 100 mL MilliQ water was heated close to 90 °C and, when cooled down to 60 °C–70 °C, gelatin was added accompanied with constant stirring. Concentration of 2 g per 100 mL and 4 g per 100 mL of gelatin in MilliQ water was used in different experiments depending on the required thickness of the nanocomposites. 150–200 μL of the prepared gelatin solution was then spin coated (Laurell Tech.,WS-650MZ23NPP/LTE) on top of the dried AgNW-Si substrates at (i) 1000 rpm at 500 rpm acceleration for 10 seconds followed by (ii) 2000 rpm at 1000 acceleration for 60 seconds. Once coated, the samples were annealed (Hotplate PSelecta,Platronic) at 100 °C for 10 minutes. The prepared samples were stored in vacuum chambers (Kartell Pvt. Ltd) until their use. For measurements under ambient conditions, the samples were subjected to ambient air for at least fifteen minutes prior to the measurements to ensure they reach a stable hydration.

**Electrostatic force volume microscopy.** Electrostatic force volume microscopy measurements have been performed using the approach detailed elsewhere,[54] by using a Nanowizard 4 BioAFM from JPK (now Bruker). Briefly, an ac voltage of frequency $f_{el}$ = 2 kHz and amplitude $v_0$ = 4 V has been applied between the conductive tip (CDT-CONTR $f_0$ = 17 kHz) and the highly doped silicon substrate by means of an external lock in (eLockin 204/2 Anfatec). Force volume data have been acquired using the Advanced Quantitative Imaging (JPK) module. At each pixel the vertical deflection and the electrical 2ω-oscillation amplitude ($A_{2\omega}$) of the cantilever have been measured as a function of tip sample distance. Typically, images of 128 × 128 pixels have been acquired. The length of the approach curve was set to ΔZ = 400 nm and each one contained 300 data points. In a post-processing step the electrical 2ω-oscillation amplitude ($A_{2\omega}$) is converted to capacitive gradient as explained elsewhere.[49] Constant height EFM images have been retrieved from the force volume data at the desired heights following the procedure described in ref. 54. Experiments were performed both under ambient conditions (relative humidity, RH ~ 40–50%) and under dry ambient conditions (RH < 10%), with the use of a custom-made environmental chamber. RH values are too small to produce a significant swelling or softening of the gelatin samples. We note that the RH can affect significantly the optical sensitivity of the instrument. To prevent tip contamination, different probes have been used for the different samples analysed in the present work. The applied electric voltage does not produce electrochemical effects in the gelatine since it is an alternating voltage of relatively high frequency and because the actual voltage drop in the gelatin sample is only a small fraction (roughly a 10%) of the applied voltage, since most of the voltage drops in the air gap between the tip and the sample.

**Finite element numerical calculations.** The quantitative analysis of the EFM approach curves acquired has been carried out following the methods of Scanning Dielectric Microscopy[49,50] adapted to deal with force volume data sets.[54] In the quantitative analysis we used the model described in Fig. 2k. In this, the tip is modelled as usual[49,50,58,59] by a truncated cone with height H and half-angle $\theta$ terminated in a tangent sphere of radius $R$. At the top of the cone a "cantilever" disk of thickness $W$ and radius $L_c = L + H.tan(\theta)$ is introduced to model local cantilever effects. The lever portion of the probe was not explicitly modelled, and its effects were included via a phenomenological capacitance gradient offset, $C'_{offset}$. The gelatin has thickness $t_m$ and dielectric constant $\varepsilon_m$, and it presents a cylindrical cap protrusion of height $h_p$ and width $w_p$. The nanowire is located at depth $d$ (measured from the upper nanowire edge to the gelatin baseline surface) and has diameter, $D_w$, length $l_w$ μm and dielectric constant $\varepsilon_w$. The tip is located at a distance $z$ from the gelatin surface. The electrostatic force acting on the tip was determined by solving the Poisson equation for the model described and integrating the Maxwell stress tensor on the tip surface, using the electrostatic module of Comsol Multiphysics 5.3 and custom codes written in

Matlab (The Mathworks) as detailed in previous works.[49,50,58,59]

**Tip geometry calibration and gelatin dielectric constant.** The tip radius, half cone angle and capacitance gradient offset used in the theoretical model were determined as detailed elsewhere[49,50] by calculating numerically capacitance gradient dC/dz approach curves for a tip-on-metal model and fitting them to the experimental *dC/dz* curves measured on a bare region of the substrate. In the analysis, the microscopic parts of the tip were adjusted to their nominal values $H = 12.5$ μm, W = 3 μm and L = 3 μm. N = 300 experimental *dC/dz* curves on the bare substrate were typically analysed. The value of the gelatin dielectric constant was determined as detailed elsewhere[50,59] by numerically calculating capacitance gradient *dC/dz* approach curves for a tip-on-thin film model and fitting them to the experimental *dC/dz* curves measured in a region of the gelatin not containing nanowires. In the calculations, the tip geometry determined earlier is used. N = 300 experimental *dC/dz* curves on the bare gelatin were typically analysed.

**Acknowledgements**

This work was partially supported by the Spanish Ministerio de Economıa, Industria y Competitividad and EU FEDER through Grant No. PID2019-111376RA-I00 and the Generalitat de Catalunya through Grant No. 2017-SGR1079, and the CERCA Program. This work also received funding from the European Commission under Grant Agreement No. H2020- MSCA-721874 (SPM2.0). R. F and L. F. received funding from the Marie Sklodowska-Curie Actions (grants 842402, Dielec2DBiomolecules) and the European Research Council (grant agreement no. 819417, Liquid2DM) under the European Union's Horizon 2020 research and innovation program. We acknowledge Dr A. Kyndiah for support in the preparation of the nanocomposite materials.

# Supplementary Information

**S1. Photodiode and tip geometry calibration and gelatin dielectric constant determination for Fig. 2**

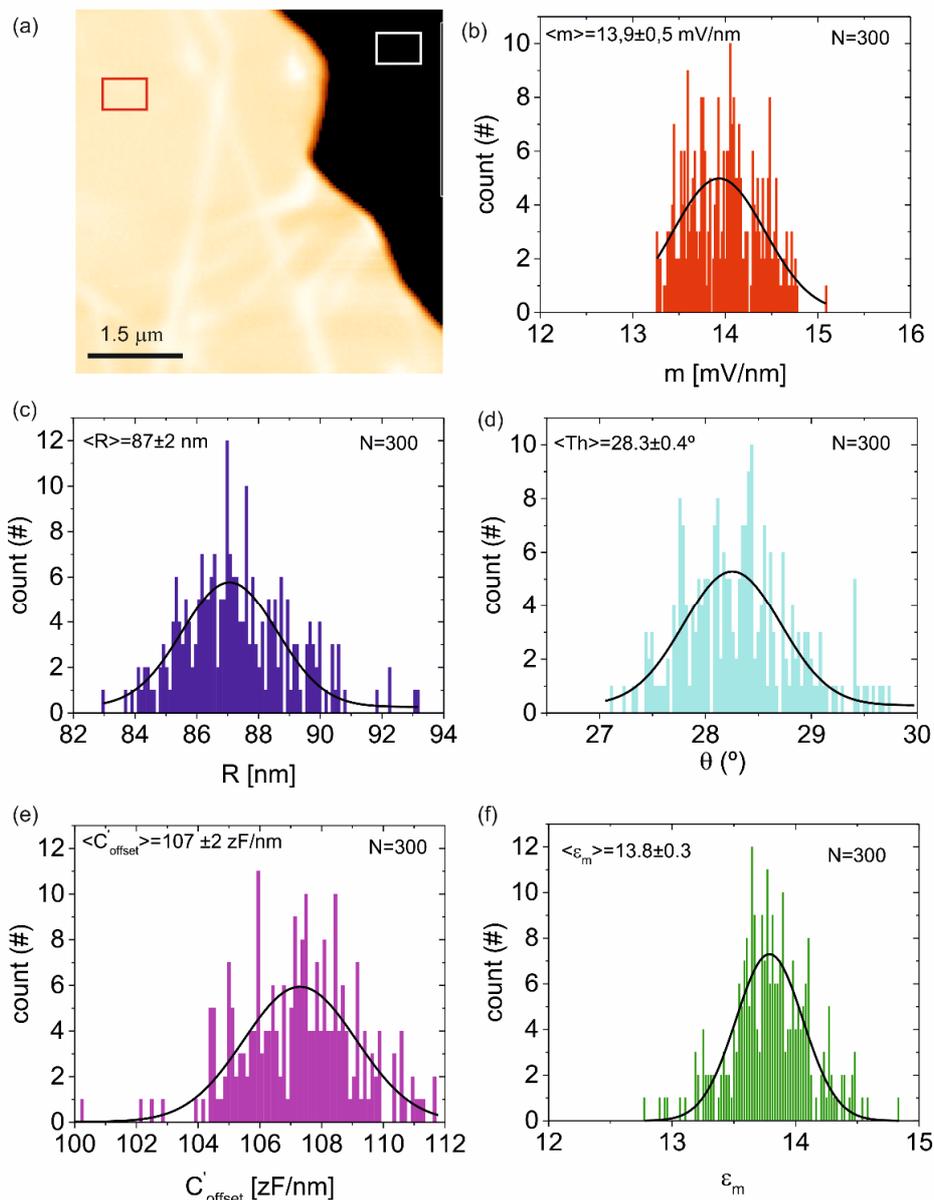

Figure S1. (a) AFM topographic image of a small region close to the scratch in Fig. 1 of the main text. The white and red rectangles highlight the regions used to calibrate the photodiode sensitivity and tip geometry and to extract the dielectric constant of gelatin, respectively. (b) Distribution of the photodiode sensitivity extracted from the slope of the contact part of the normal deflection approach curves. The mean value (N=300) is 13.9±0.5 mV/nm. (c)-(e) Histogram of the tip radius, half cone angle and capacitance gradient offset obtained by fitting theoretical $dC/dz$ curves generated for a tip-on-metal model to the experimental $dC/dz$ approach curves acquired in the white rectangle region in (a). The mean values (N=300) obtained are $R$=87±2 nm, θ=28.3±0.4º and $C'_{offset}$=107±2 zF/nm, respectively. (f) Histogram distribution of the values obtained for the dielectric constant of gelatin obtained by fitting theoretical $dC/dz$ curves generated for a tip-on thin dielectric film model to the experimental $dC/dz$ approach curves acquired in red rectangle in (a). The mean value (N=300) obtained is $\varepsilon_m$=13.8±0.3.

**S2. Effect of the nanowire diameter in the estimation of the depths.**

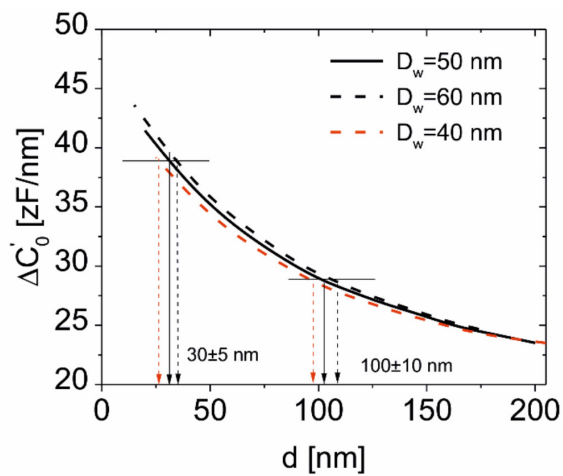

Figure S2. Capacitance gradient contrast at the center of a buried nanowire, $\Delta C'_0$, calculated with the model in Fig. 2k of the main text as a function of the depth, for three nanowire diameters $D_w$=40 nm, 50 nm and 60 nm and a surface protrusion $h_p$ = 8 nm. The tip is located at a height $z$ = 36 nm from the flat surface. From this graph we can estimate that a variation of ± 10 nm of the nanowire diameter of 50 nm induces at most an error of ±5 nm in the estimation of the depth. Parameters: same as those in Fig. 2.

## S3. Effect of the length of the nanowire in the estimation of the depth

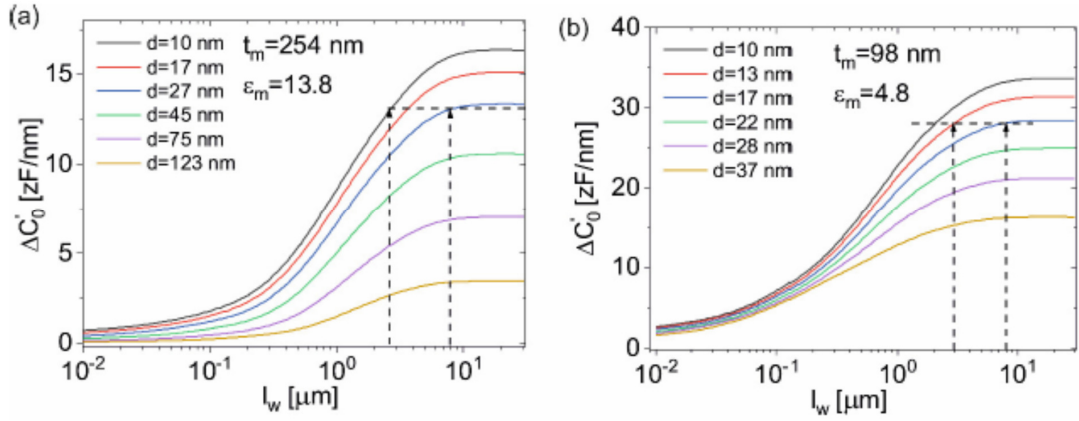

Figure S3. (a) Capacitance gradient contrast at the center of a buried nanowire for the tip at a distance $z = 36$ nm from the surface of the gelatin film as a function of the length of the nanowire, for different depths, $d$, with no surface $h_p = 0$ nm. Parameters of the calculations: Same as for Fig. 2 of the main text. For lengths larger than $l_w \sim 8$ μm (the one used in the main text) the results become independent from the length. If a nanowire is shorter, one can estimate the depth as indicated by the arrows in the graph. (b) Idem for the case of parameters corresponding to Fig. 4 of the main text.

## S4. Inclined vs non-inclined nanowire models

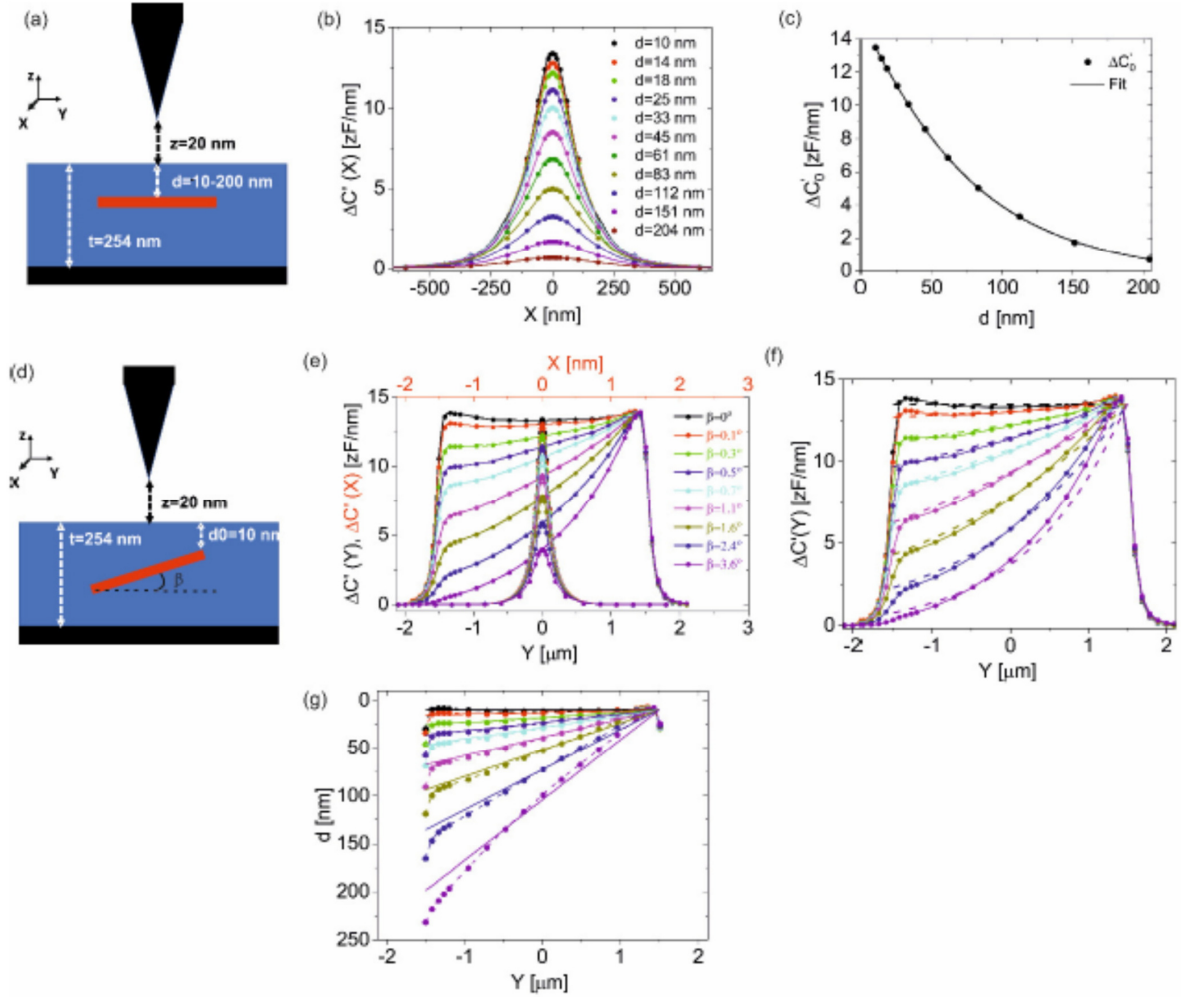

Figure S4. (a) Schematic representation of a model of a buried nanowire parallel to the substrate (same model as the one used in the main text, without the surface protrusion). (b) Cross-section capacitance gradient contrast profiles along the transversal direction to the nanowire, ΔC'(X), for different depths, $d$. (c) (symbols) Capacitance gradient contrast at the center of the wire, ΔC'$_0$, as a function of the depth. (continuous line) Least square fitting of the function $\Delta C'_0(d) = A(1+Bd)/(1+Cd+Dd^2)$, with A=15.1 zF/nm, B=-3.7·10$^{-3}$nm$^{-1}$, C=7.1·10$^{-3}$nm$^{-1}$, D=7.0·10$^{-5}$nm$^{-2}$. (d) Schematic representation of a model of a buried nanowire inclined an angle β with respect to the substrate. (e) Capacitance gradient contrast profiles along the transversal, ΔC'(X), (black symbols, left and bottom axes) and longitudinal, ΔC'(Y), (red symbols, right and top axis) directions, respectively. (f) (continuous lines) Capacitance gradient contrast profile along the nanowire, ΔC'(Y), calculated by using the function ΔC'$_0$(d) and the depth profile of inclined nanowires, $d(Y) = d_0 + [(l_w/2)\cos(\beta) - Y]\tan(\beta)$. The symbols correspond to the numerically calculated profiles (same as in (e)). (g) (symbols) Local nanowire depths extracted with the function ΔC'$_0$(d) applied to the profiles of the inclined nanowires in (e), ΔC'(Y). The extracted depths nicely reproduce the actual nanowire depth profiles (continuous lines). This result demonstrates that the non-inclined nanowire model can be used to predict the local depth of inclined nanowires, as we did in the manuscript. Parameters used in the calculations: $D_w$=50 nm, $l_w$=3 μm, $\varepsilon_w$=10$^5$ (metallic), $d_0$=10 nm, $t_m$=254 nm, $\varepsilon_m$=13.8, $R$=87 nm, θ=28°, $H$=12.5 μm, $W$=3 μm, $L$=3 μm.

## S5. Photodiode and tip geometry calibration, and gelatin dielectric constant determination for Fig.4

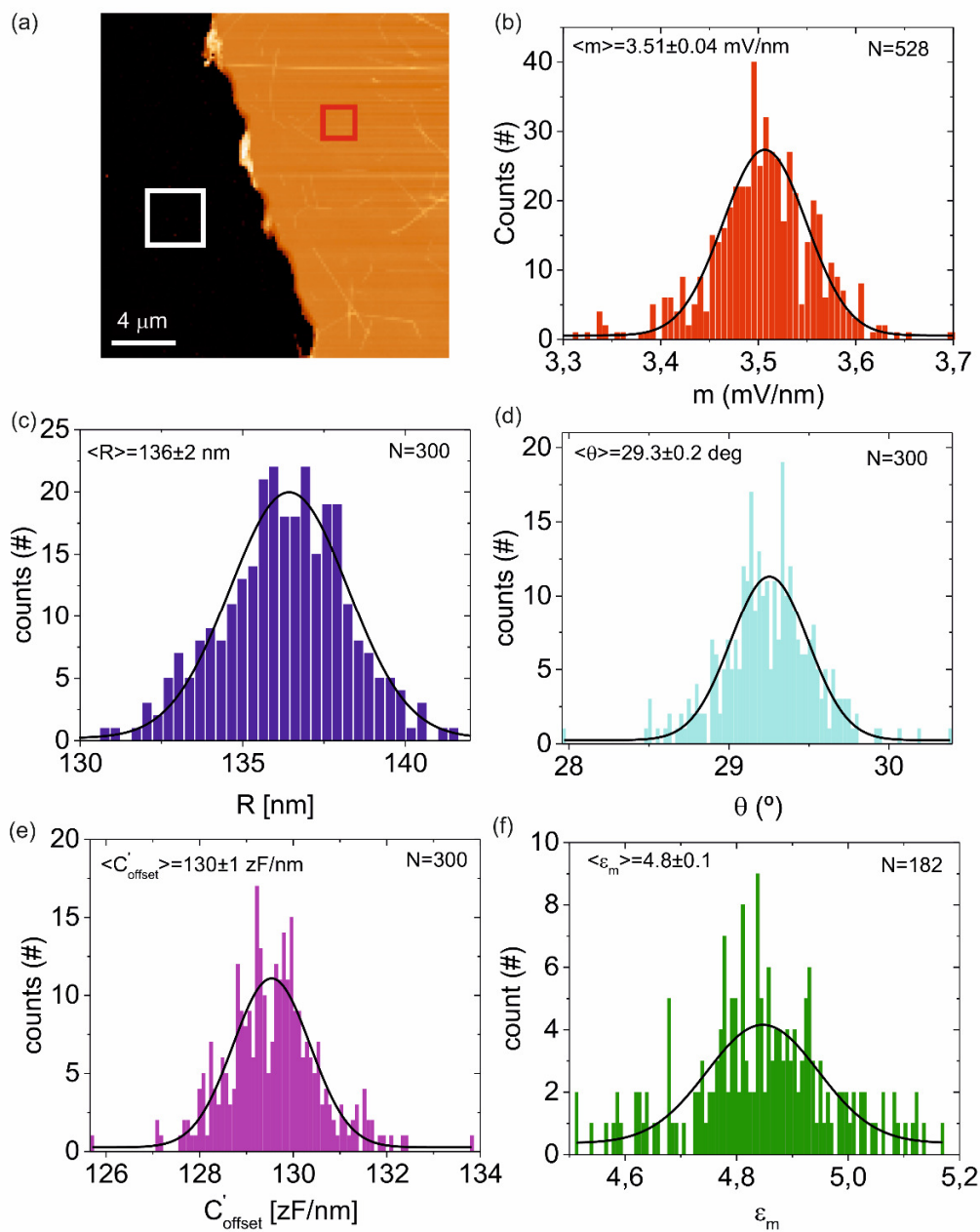

Figure S5. (a) AFM topographic image of a larger region corresponding to Fig. 4 of the main text. The white and red rectangles highlight the regions used to calibrate the photodiode sensitivity and tip geometry and to extract the dielectric constant of gelatin, respectively. (b) Distribution of the photodiode sensitivity extracted from the slope of the contact part of the normal deflection approach curves. The mean value (N=528) is 3.51 ± 0.04 mV/nm. (c)-(e) Histogram of the tip radius, half cone angle and capacitance gradient offset obtained by fitting theoretical $dC/dz$ curves generated for a tip-on-metal model to the experimental $dC/dz$ approach curves acquired in the white rectangle region in (a). The mean values (N=300) obtained are $R$ = 136±2 nm, $\theta$ = 29.3±0.2° and $C'_{offset}$ = 130±1 zF/nm, respectively. (f) Histogram distribution of the values obtained for the dielectric constant of gelatin obtained by fitting theoretical $dC/dz$ curves generated for a tip-on thin dielectric film model to the experimental $dC/dz$ approach curves acquired in red rectangle in (a). The mean value (N=300) obtained is $\varepsilon_m$=4.8±0.1

**S6. Examples of experimental capacitance gradient approach curves and of the corresponding fitted curves**

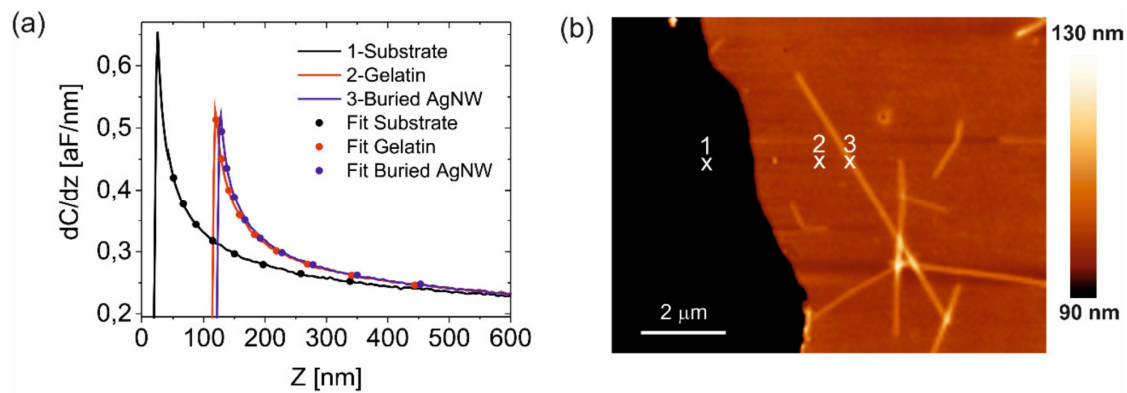

Figure S6. (a) (continuous lines) Examples of experimental capacitance gradient approach curves, *dC/dz*, corresponding to three pixels of the image in Fig. 4 of the main manuscript at the positions indicated by the crosses in (b). The symbols represent the corresponding least square fitted theoretical curves. The theoretical curve on the substrate is generated from a tip-flat metal model, that on the gelatin from a tip-thin film dielectric model and that on the buried nanowire from the model in Fig. 2k of the main manuscript. From the fit on the substrate curve one obtains the tip geometry, from that on the gelatin the dielectric constant of gelatin and from that on the buried nanowire the depth position